\documentstyle[iopconf1,epsf]{article}
\newcommand{\be}{\begin{equation}}
\newcommand{\ee}{\end{equation}}
\newcommand{\hmp}{h^{-1}Mpc}
\newcommand{\bef}{\begin{figure}}
\newcommand{\eef}{\end{figure}}
  
\def\spose#1{\hbox to 0pt{#1\hss}} 
\def\ltapprox{\mathrel{\spose{\lower 3pt\hbox{$\mathchar"218$}} 
 \raise 2.0pt\hbox{$\mathchar"13C$}}} 
\def\gtapprox{\mathrel{\spose{\lower 3pt\hbox{$\mathchar"218$}} 
 \raise 2.0pt\hbox{$\mathchar"13E$}}} 
\def\inapprox{\mathrel{\spose{\lower 3pt\hbox{$\mathchar"218$}} 
 \raise 2.0pt\hbox{$\mathchar"232$}}}

\begin{document}
\title{On the fractal structure of the universe: 
methods, results and theoretical implication}

\author{Francesco Sylos Labini\footnote{E-mail:
sylos@amorgos.unige.ch}}

\affil{D\'ept.~de Physique Th\'eorique, Universit\'e de Gen\`eve,  
		24, Quai E. Ansermet, CH-1211 Gen\`eve, Switzerland
		and INFM Sezione Roma1 P.le A. Moro, 2,  
        	I-00185 Roma, Italy.}

\beginabstract
The fact that galaxy distribution
exhibits fractal properties is   well
established since twenty years.
Nowadays, the controversy concerns 
the range of the fractal regime,
the value of the fractal dimension and
the eventual presence of a cross-over
to homogeneity. Fractal properties
maybe studied with methods 
which do not assume homogeneity a priori
as the standard statistical methods do.
We show that complementary 
to the adoption of   new methods of analysis
there are important theoretical implications
for the usual scenario of galaxy formation.
For example, we focus on the concept of bias 
and we show that it needs
a basic revision 
even if  future redshift surveys
will be able to identify an eventual
tendency to homogenization.
\endabstract

\section{Introduction}

 The assumption of homogeneity in the distribution of matter
lies at the heart of the Big Bang cosmology. The
nature of the evidence, if any, for this assumption 
has, however, been the subject of very considerable controversy
\cite{dav97,pmsl97}. 
A central point made by Pietronero \cite{pie87} has been 
that the standard methods of analysis of galaxy red-shift 
catalogues, which provide the most direct probe of 
the (luminous) matter distribution,
actually assume homogeneity implicitly. 
In this report we review the main 
points of this controversy: we discuss the different 
methods
of analysis of redshift samples, and the
corresponding  results. The basic point 
we try to clarify is: what do we learn from the 
redshift surveys ? We show that complementary 
to the adoption of a new method of analysis
there are important theoretical implications
for the usual scenario of galaxy formation.
Moreover we briefly discuss the results
of N-body simulations. We refer to
\cite{slmp98} for a more detailed and 
complete picture.

\section{Redshift surveys and the usual methods of analysis}

Until the seventies galaxy distributions were only known in terms of 
angular catalogues.
These catalogues are 
limited by the  apparent luminosity of galaxies. 
Since the intrinsic luminosity can vary over a range of the order 
of one million, the points of the angular catalogues 
correspond to extremely 
different distances and they are a complex 
convolution of these. These angular  
distributions appeared rather smooth and, as such, 
they  
justified the usual statistical assumptions 
of homogeneity at large angular separations.
Actually the measurement of the {\it amplitude}
of angular fluctuations with respect to the
(angular) average density shows that 
at large enough scale 
galaxy distribution turns out to
be very smooth and uniform \cite{pee80,pee93}.

 The redshift measurements permit to locate galaxies  
in 3-d space.
 They immediately showed a clumpy distribution with large clusters  
and large voids, in apparent contrast with the angular data.  
This finally led to 3-d catalogues from which one could make volume limited 
samples (i.e. samples which are not biased by 
any observational luminosity selection effect \cite{slmp98}), 
which provide the best and most direct  
information  
for the correlation analysis. These 3-d data have been and are extensively 
analysed with the usual $\xi(r)$ method.  
By using the standard two-points correlation function
 it has been found \cite{pee93,dav97}
that galaxy structures are characterized 
by having a very well defined "correlation length"
that is found to be $r_0 \approx 5 \hmp$.
The physical interpretation of such a scale
being the distance above which
the density fluctuations become
of the same order of the average density. At twice this
distance the fluctuations have
small amplitude and the linear theory holds,
i.e the distribution becomes homogeneous.

Most authors who analyse galaxy (and cluster) catalogs use 
these methods, and they are the instruments in terms of 
which the predictions of all standard type theories 
of structure formation are framed.  That the ``correlation length'' 
is a real physical scale characterizing the clustering of
galaxies is affirmed by producing evidence for
the stability of this scale in different samples.
In practice, however, correlation lengths are not
observed to be very stable, and it is here that the
concept of ``luminosity bias'' enters \cite{dav88,par94,ben96}: Galaxies 
of different brightness are supposed to be clustered 
differently, and it is this which is proposed as
the physical explanation of the observed variation,
rather than the absence of real underlying homogeneity. 
Rather than their being one real scale characterizing 
the correlations of galaxies, there is then an
undetermined number of such scales. 
The luminosity bias effect is therefore the cornerstone
of the concept of bias: In the standard biasing picture the 
distribution of dark matter, galaxies  and galaxy clusters
are  Gaussian but 
with different {\it amplitudes}, i.e. different correlation lengths.  

In summary, the most popular point of view, is   that
galaxy distribution
exhibits indeed fractal properties
with dimension $D \approx 1.3$ at  small
scale (i.e. up to $\sim 10 \div 20  \hmp$).
The main result of this approach is that a characteristic length  
is derived $r_0 \approx 5 \hmp$ which should  
mark the tendency towards homogenization. This is in apparent 
agreement with the structureless angular data but it is puzzling  
with respect to the structures of the 3-d data. 
From this perspective it seems that the presence or absence  
of structures  
in the data is irrelevant for the determination of 
$r_0$.  
 
Puzzled by these results we
decided to reconsider the 
question of correlations 
 from a broader and critical perspective \cite{cp92,slmp98}. 
This allowed us to test homogeneity, 
instead of assuming it as in the $\xi(r)$ analysis, 
and to produce a totally unbiased description of the correlation properties  
using the methods of modern Statistical Physics.  
In the next section we discuss the methods and
we show the main results.
Very recently a wide debate on this subject is
in progress \cite{gu97,slmp98b,coles98,sca98,joyce198,tee98,
cappi98,joyce298,wu98,pee98,man98,psl98}
\footnote{See the web page
http://www.phys.uniroma1.it/DOCS/PIL/pil.html
where all these materials have been collected}.

\section{Correlation properties of redshift samples }
 
An alternative analysis has been proposed using statistical methods 
which  do not make the  assumption of homogeneity, and are appropriate
for characterizing the properties of regular as well as irregular
distributions \cite{pie87,cp92,slmp98}.
Essentially this analysis makes use of
a very simple statistic, the conditional
average density, defined as
\be 
\label{e4} 
\Gamma(r) = \left \langle \frac{1}{S(r)} \frac{dN(<r)}{dr} \right \rangle 
\ee 
where $dN(<r)$ is the number of points in a shell 
of radius $dr$ at distance $r$ from an occupied point and 
$S(r)dr$ is the volume of the shell. Note that in Eq.\ref{e4} 
there is an average over all the occupied points contained in the sample. 
If the distribution is scale invariant with fractal dimension $D$
we have
\be 
\label{ee5} 
\Gamma(r) \sim r^{-\gamma} \;  
\ee 
where $\gamma=3-D$. Clearly the conditional 
average density is an appropriate statistical tool 
with which to study fractal versus homogeneous properties 
in a given distribution, and in particular the approach 
(if any) to homogeneity. The problems of the standard
analysis can easily be seen from the fact that, for
the case of a fractal distribution
the standard ``correlation function'' $\xi(r)$ 
in a spherical sample of radius $R_s$, 
is given by 
\be 
\label{e7} 
\xi(r) = \frac{3-\gamma}{3}  
\left( \frac{r}{R_s} \right)^{-\gamma} -1 \; . 
\ee 
Hence for a fractal structure  the ``correlation length'' $r_0$ 
(defined by $\xi(r_0)=1$), is not a scale characterizing any
intrinsic property of the distribution, but just a scale
related to the size of the sample. If, on the other hand, the
distribution is fractal up to some scale $\lambda_0$ and
homogeneous beyond this scale, it is simple to show that
(if $\lambda_0 < R_s$, i.e. the crossover to homogeneity 
is well inside the sample size)  
\be 
\label{e9} 
r_0 = \lambda_0 \cdot 2^{-\frac{1}{\gamma}} \; . 
\ee 
The correlation length does in this case have a real physical meaning
(when measured in samples larger than $\lambda_0$),
being related in a simple way to the scale characterizing
homogeneity. In the case $D=2$ we have $r_0 = \lambda_0/2$
\footnote{This calculation assumes a simple matching of
a fractal onto a pure homogeneous distribution. For 
any particular model with fluctuations away from 
perfect homogeneity, the numerical factor will differ
slightly depending on how precisely we define the 
scale $\lambda_0$.}.
Finally it should be noticed that 
$\:\xi(r)$ is power law only for
\be
\label{xi3}
\left(\frac{3-\gamma}{3}\right) 
\left(\frac{r}{R_{s}}\right)^{-\gamma}  \gg 1
\ee
hence for $r \ll r_0$: for larger distances there is a clear deviation
from the power law behavior due to the definition of $\xi(r)$.
This deviation, however, is just due to the size of
 the observational sample and does not correspond to any real change
of the correlation properties. It is clear that if one 
estimates the  $\xi(r)$ 
exponent  at distances $r \ltapprox r_0$, one 
systematically obtains a higher value of the correlation exponent
due to the break of $\xi(r)$ in the log-log plot. 
In this respect it is useful to compute
the log derivative of eq.\ref{e7} with respect to $\log(r)$:
\be
\label{xi4}
\gamma'=\frac{d(\log(\xi(r))}{d\log(r)} = - \frac{2 \gamma r_0^{\gamma} r^{-\gamma}} 
{2 r_0^{\gamma} r^{-\gamma} -1}
\ee
where $r_0$ is defined by $\xi(r_0) = 1$.
 The tangent
to $\xi(r)$ at $r=r_0$ has a slope $\gamma'=-2\gamma$.
It is clear   that, even if the
distribution has fractal properties, it is very difficult
to recover the correct slope from the study of 
the $\xi(r)$ function. The $\xi(r)$ is intrinsically
problematic to this end.

\begin{figure}
\epsfxsize 11cm
\centerline{\epsfbox{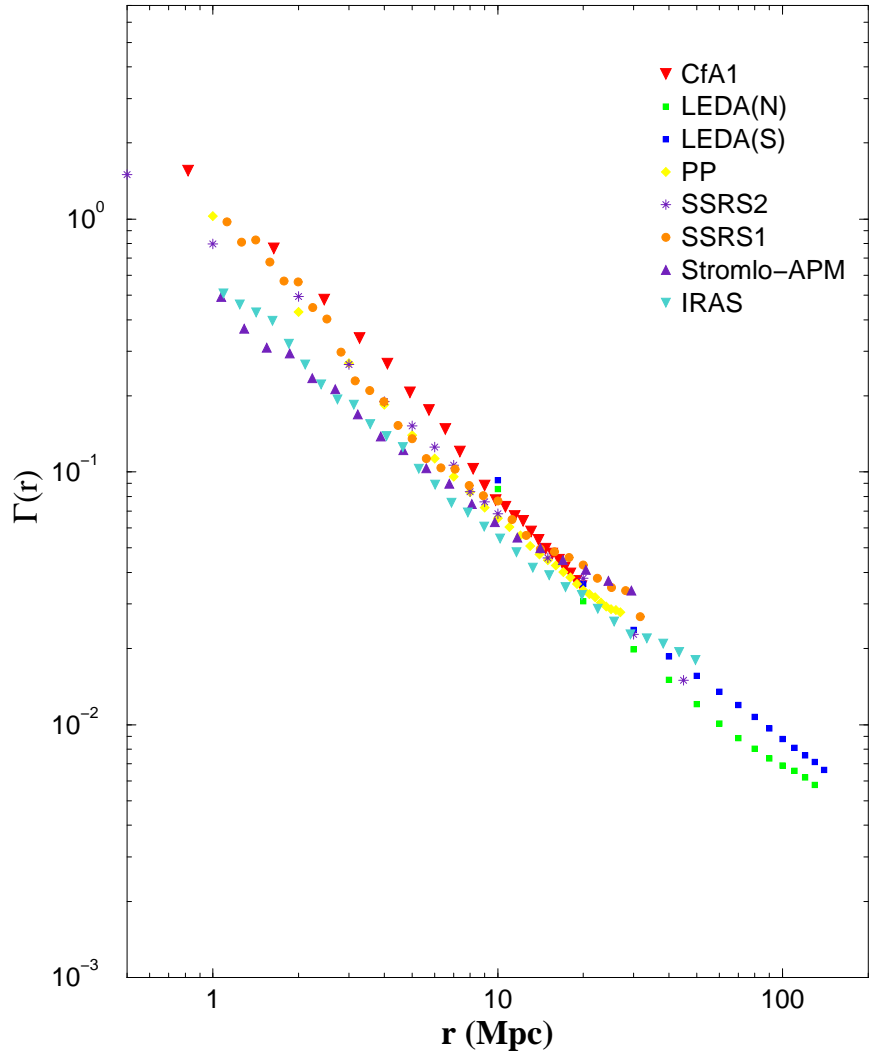}} 
\caption{Full correlation analysis for the various
 available redshift surveys in the range of distance $\sim 1 \div 200
 \hmp$. A reference line with slope $-1$ is also shown, which 
corresponds to fractal dimension $D = 2$. \label{gamma}}
\end{figure}

The direct analysis of all available
galaxy catalogues reported in \cite{slmp98}, using the 
conditional average density $\Gamma(r)$.
 In Tab.\ref{tab1} we report the characteristics of the 
 various catalogs we have analyzed by using
 the methods previously illustrated.
 \begin{table}
 \caption{\label{tab1}The volume 
 limited catalogues are characterized by the following 
 parameters:
 - $R_d (\hmp)$ is the depth of the catalogue
 - $\Omega$ is the solid angle
 - $R_s (\hmp)$ is the radius of the largest sphere 
 that can be contained in the catalogue volume. 
 This gives the limit of statistical validity of the sample.
 - $r_0(\hmp)$  is the length at which $\xi(r) \equiv 1$.
 - $\lambda_0$ is the eventual real crossover to a homogeneous 
 distribution that is actually never observed. 
 The value of $r_0$ 
 is the one obtained 
 in the deepest VL  sample. 
 (Distances are expressed in $\hmp$).
 }
 \begin{tabular}{|c|c|c|c|c|c|c|}
 \hline
     &      &          &    &              &  &       \\
 \rm{Sample} & $\Omega$ ($sr$) & $R_d  $ & $R_s  $ 
 & $r_0  $& $D$ & $\lambda_0 $ \\
     &       &    &    &    &              &       \\
\hline
CfA1         & 1.83      & 80 & 20 & 6  & $1.8 \pm 0.2$ & $ >80 $    \\
CfA2South    & 1.23      & 130& 50 & 15 & $2.0 \pm 0.1$ & $ >120$    \\
PP           & 0.9       & 130& 30 & 10 & $2.0 \pm 0.1$ & $ >130$    \\
SSRS1        & 1.75      & 120& 35 & 12 & $2.0 \pm 0.1$ & $ >120$    \\
SSRS2        & 1.13      & 150& 45 & 15 & $2.0 \pm 0.1$ & $ >130$    \\
Stromlo-APM  & 1.3       & 100& 35 & 12 & $2.2 \pm 0.1$ & $ >150$    \\
LEDA         & $\sim 5 $ & 300& 150& 45 & $2.1 \pm 0.2$ & $ >150$    \\
LCRS         & 0.12      & 500& 18 & 6  & $1.8 \pm 0.2$ & $ >500$    \\
IRAS$2 Jy$   & $\sim 5$  & 60 & 20 & 5  & $2.0 \pm 0.1$ & $ >50$     \\
IRAS$1.2 Jy$ & $\sim 5$  & 80 & 30 & 8  & $2.0 \pm 0.1$ & $ >50$     \\
ESP          & 0.006     & 700& 8  & 3  & $2.0 \pm 0.2$ & $ >700$    \\
             &           &         &    &               &    &       \\
\hline
\end{tabular}
\end{table}
 We show in Fig.\ref{gamma} the results
 of the conditional density
 determinations in various redshift surveys \cite{slmp98}.
 All the available data 
 are consistent with each other and show fractal correlations with
 dimension $D = 2.0 \pm 0.2$ up to the deepest scale probed up to
 now by the available redshift surveys, i.e. $\sim 150 \hmp$.
 A similar result has been obtained by the analysis of galaxy cluster
 catalogs \cite{slmp98}. If we consider also the determination
 of the radial density, which is a much weaker test 
 because it does not involve an average \cite{slmp98,joyce198}, 
 we find  a rather good continuation of 
 the fractal behavior up to $\sim 1000 \hmp$.

\begin{figure}
\epsfxsize 9cm
\centerline{\epsfbox{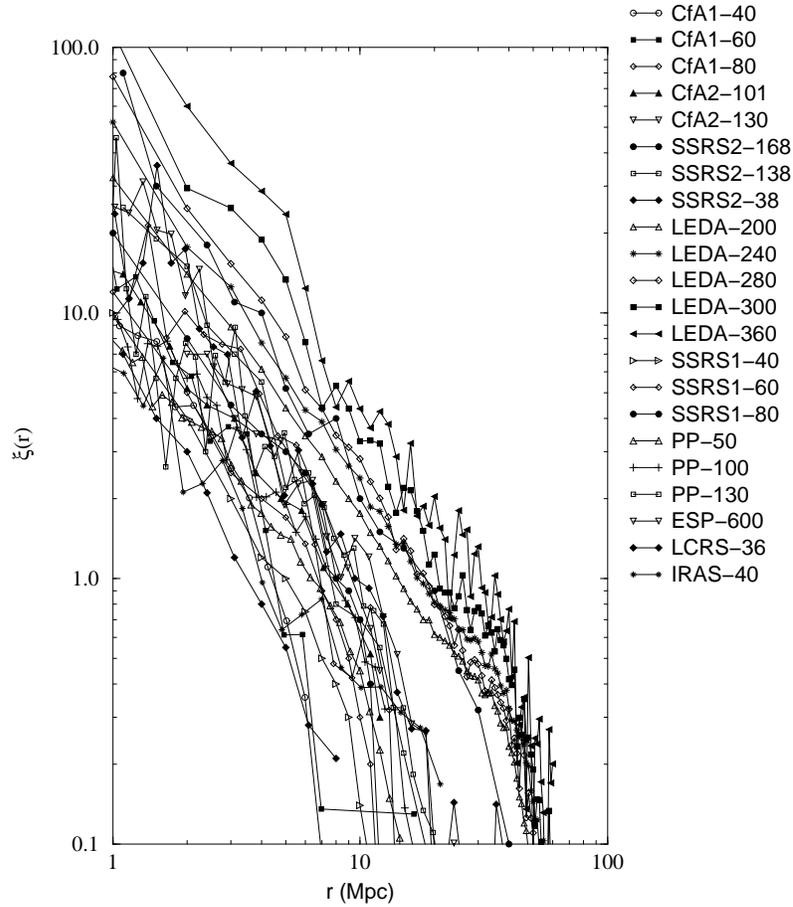}} 
\caption{\label{xi} Traditional analyses based on the function $\xi(r)$
of the same galaxy catalogs of the previous figure.
 The usual 
analysis is based on the a priori untested assumptions of 
analyticity and homogeneity. These properties
are not present in the real galaxy distribution and 
the results appear therefore rather confusing. 
This lead to the impression that galaxy catalogs are not good
enough and to a variety of theoretical problems like the 
galaxy-cluster mismatch, luminosity segregation, linear and 
non-linear evolution, etc. This situation changes completely and 
becomes quite clear if one adopts the more 
general conceptual framework that is at the basis 
the previous figure}
\end{figure}

It is interesting to compare the analysis of Fig.\ref{gamma} with 
the usual one, made by the function $\xi(r)$, for the same 
galaxy catalogs. This is reported in Fig.\ref{xi}.
 From this point of view, the various data 
the various data appear to be in strong disagreement with 
each other. This is due to the fact that the usual analysis
looks at the data from the prospective of analyticity and large
scale homogeneity (within each sample). These properties have never
been tested and they are not present in the real galaxy
distribution so the result is rather confusing (Fig.\ref{xi}).
Once the same data are analyzed with a broader perspective the
situation becomes clear (Fig.\ref{gamma}) and the data of 
different catalogs result in agreement with each other. It is 
important to remark that analyses like those of Fig.\ref{xi}
have had a profound influence in the field in various ways: 
first the different  catalogs appear in conflict with each other.
This has generated the concept of {\it not fair samples} and a 
strong mutual criticism about the validity of the data 
between different authors. In the other cases the
discrepancy observed in Fig.\ref{xi} have been 
considered real physical problems for which various technical
approaches have been proposed. These problems are, for example, 
the galaxy-cluster mismatch, luminosity segregation, 
the richness-clustering relation and 
 the linear non-linear evolution of the perturbations 
corresponding to the {\it "small"} or  {\it "large"}
amplitudes of fluctuations. All this problematic
situation {\it is not real} and it arises only from a 
statistical analysis based on 
inappropriate and too restrictive
 assumptions that do not find 
 any correspondence in the 
physical reality. It is also
important to note that,
even if the galaxy distribution 
would eventually became 
homogeneous at larger scales, the use of the above statistical
concepts  is anyhow inappropriate for the range of scales 
in which the system shows fractal correlations as those 
shown in Fig.\ref{gamma}.

\subsection{Other length scales}

The usual analysis finds that 
 rms fluctuations of the observed galaxy density field are very large 
on small scales, of order of unity within spheres of $8 \hmp$ dropping as a
power law as a function of scale, becoming few percent
at several tens $\hmp$. In this perspective it therefore makes sense 
to reference to the density field of galaxies to its mean.
Let $\rho(r)$ be the observed 
galaxy density field; the density fluctuation
field is defined as
\be
\label{d1}
\delta(r) = \frac{\rho(r) - \langle \rho \rangle}{\langle \rho \rangle}
\ee
This quantity can be measured in redshift samples. 
The problem in this case is the same 
one which enters in the definition of $r_0$:
one is comparing the amplitude of fluctuations
to the mean density. As an example, 
one  can consider a
portion of a fractal structure
of size $\:R_{s}$ and study
the behavior of  $\:\delta N / N$.
The average density   is
just given by Eq.\ref{ee5} while
the overdensity
$\:\delta N$, as a function
of the size $\:r$  ($\:r\leq R_{s}$) of a given
in structure is:
\be
\label {e3n1}
\delta N = \frac{N(r)}{V(r)} - <n>
= \frac {3}{4\pi} B (r^{-(3-D)}-R_{s}^{-(3 - D)}) \; .
\ee
We have therefore
\be
\label {e3n2}
\frac {\delta N}{N} =
\left(\frac{r}{R_{s}}\right)
^{-(3 - D)} - 1 \; .
\ee
Clearly for structures that
approach the size of the 
sample, the value of  $\:\delta N / N$
becomes very small and eventually
becomes zero at $\:r = R_{s}$.

Another typical length scale which 
is usually defined in the study of redshift samples 
is the scale at which the power spectrum (hereafter PS),
of the {\it density fluctuations}
has a turnover: $dP(k(\lambda_{f}))/dk = 0$. 
Essentially all the currently
elaborated models of galaxy formation 
(e.g \cite{pee93})
{\it assume large scale homogeneity} and 
predict that the galaxy
PS, 
which  is {\it the PS  of the density contrast},
decreases both toward small scales and toward large
scales, with a turnaround somewhere in the middle, at a scale $\lambda_f$
that can be taken as separating ``small'' from ``large'' scales. 
Because of the homogeneity assumption, the PS    amplitude
should be independent on the survey scale, any  residual 
variation being attributed to luminosity bias (or to the
fact that the survey scale has not yet reached the homogeneity scale).
However, the crucial clue to this
picture, the firm determination of the 
scale $\lambda_f$, is still missing, although
some surveys do indeed produce a  turnaround
 scale around 100 $\hmp$. 
Recently, the CfA2  survey 
analyzed by \cite{par94} (PVGH), showed a $n=-2$ slope up to $\sim 30 \hmp$,
a milder $n\approx -1$ slope  up to 200 $\hmp$, and some tentative
indication of flattening on even larger scales. PVGH also find
that deeper subsamples have higher power amplitude,
i.e. that the amplitude scales with the sample depth.

It is simple to show \cite{slmp98} 
that both features, bending and scaling,
are a manifestation of 
the finiteness of the survey volume, and that they
cannot be
interpreted as the convergence to homogeneity, nor to a PS   
flattening. 
The systematic effect of the survey finite size is in fact
to suppress
power at large scale, mimicking a real flattening.
In fact we have shown  that
even a fractal distribution of matter, 
which never reaches  homogeneity, shows a sharp flattening
and then a turnaround.
 In particular, it is possible to show that 
 in a spherical sample of radius $R_s$, 
 which
 contains a portion of a fractal structure with dimension $D=2$,
 the PS turnover scale is given by
 \be
 \label{ps1}
 \lambda_f \approx 1.45 R_s
 \ee
 and hence it is another quantity related to the sample
 size rather than being an intrinsic characteristic scale
 of galaxy distribution.

\section{What do we learn from galaxy catalogs ?} 

As we have already mentioned the usual concept
pf bias arises from the interpretation 
of the results of the $\xi(r)$ analysis. 
The concept of bias fixes the relative
distribution of galaxies of different
types, clusters and dark matter.
In general \cite{strauss98} one assumes that
it exists a direct proportionality between the density fluctuations 
of galaxies $\delta_g$ and dark matter $\delta_{DM}$ 
\be
\label{bias1}
\delta_g = b \delta_{DM}
\ee
and the same concepts applies to galaxies of different
masses and galaxy clusters. Under this assumption, the biasing parameter
{\it b}  is independent of location.
One case use the two-point correlation function $\xi_{gg}(r)$
and the mass autocorrelation function $\xi_{\rho\rho}(r)$
to define the bias factor
\be
\label{bias2}
 b = \left(\frac{\xi_{gg}(r)}{\xi_{\rho\rho}(r)}\right)^{1/2} \;.
\ee

Let us see in more detail the origin of the concept of
bias as given in eq.\ref{bias2}. 
It is a well known observational fact that
galaxies of different morphological types
have different clustering properties. For example,
the  most luminous  elliptical galaxies  
usually reside in the clusters  cores, at local density maxima,  
and are not present in low density fields,  so that these objects 
seem to be the product of dense environments. 
There are various other morphological facts
of this type \cite{slmp98} which support 
the fact that brighter (more massive) galaxies 
are more {\it clustered} than for example spirals
(less massive). The different of clustering properties
has been interpreted, through the $\xi(r)$
analysis, as a different {\it amplitude of correlation}
for different galaxy types. In particular while
for the general galaxy field the correlation
length is $r_0 \approx 5 \hmp$, for the 
brighter galaxies ($L > L^*$) it has been found
\cite{par94,ben96} that $r_0 \approx 16 \hmp$.
This trend seems to be confirmed 
also by the cluster (more massive than galaxies) 
distribution for which 
$r_0 \approx 25 \hmp$\cite{bs88}.

On the contrary from our interpretation it follows 
a number of important implications in this respect.
As a crossover to homogeneity has not
been found, {\it all} the length scales
found by the $\xi(r)$ analysis are artifact of
an inconsistent data analysis. The "correlation
lengths"  $r_0 = 5, 16, 25, ... \hmp$
are not real physical characteristic scales,
but just fractions of   sample sizes.
 Brighter objects allow one  to investigate a
larger volume of space. Hence, for example
the sample size of cluster catalogs
is usually larger than the one of galaxy
samples. This simple observation explains
why one obtains different correlations lengths,
and in general why the correlation length
seems to increase as a function of the 
luminosity of objects. To this
qualitative observation, one may 
add a detailed study of the available
galaxy and cluster samples. 
This has been done in a detailed way
by our group and we refer
to \cite{slmp98} for an exhaustive
explanations of this fact.

 Therefore, contrary to the usual interpretation,
 we have shown  that the segregation of giant
 galaxies in clusters arises as a consequence of self-similarity of
 matter  distribution, and that in this case the only relevant
 parameter is the {\it exponent}  of the correlation function,
 while the amplitude is a spurious quantity that has no direct
 physical meaning and depends explicitly on the  sample size.  
 Let us explain in more detail this important point.

 In a well-known review on the galaxy luminosity  function (LF)
 Binggeli \etal  \cite{bin88} state that {\em "as the distribution
 of galaxies  is
 known to be inhomogeneous on all scales up to a least $100 h^{-1}
 Mpc$, a rich cluster  of galaxies is like a Matterhorn in a grand
 Alpine landscape of  mountain ridges and valleys of length up to
 100 Km"}. This point of
 view  can be seen in the light  of the concept of multifractality of the mass
 distribution. The main observational aspects of galaxy
 luminosity and space distributions are strongly related and can be
 naturally linked and explained  as a multifractal (MF)
 distribution. The concept of MF is appropriate to discuss physical
 systems with local  properties of self-similarity, in which the
 scaling properties are  defined by a continuous distribution of
 exponents. Roughly speaking one can visualize this property as
 having different scaling properties for different regions of the
 system (see \cite{cp92,slp96} for a more detailed discussion).
 The fundamental point which is dismissed in the usual
 picture,  is that {\it the whole matter
 distribution, i.e.  weighing each point by its mass, is 
 self-similar}.  This situation  requires the generalization of the
 simple fractal scaling  to a MF distribution in which a  
 continuous set of exponents is necessary to  describe the spatial
 scaling  of peaks of different weight  (mass or luminosity). {\it In
 this respect the mass and space  distributions become naturally
 entangled with each other.  }
 
 The MF implies a strong correlation between
 spatial and mass distribution so that the number of objects with
 mass $\:M$ in the point $\:\vec{r}$  per unit volume
 $\:\nu(M,\vec{r})$, is a function of space and mass and {\it it is
 not separable in a space density multiplied by a mass (or
 luminosity) function} \cite{bin88}.
 This
 means that  we {\it cannot} express the number of galaxies
 $\:\nu(M,x,y,z)$  lying in volume $\:dV$ at {\it (x,y,z)} with
 mass between {\it M} and {\it M + dM} as the {\it product}
 of a space density and a luminosity function. 
 This would assume 
 that galaxy positions are not correlated with their 
 luminosities, while the observations show
 just the opposite. 
  Moreover  we
 cannot define a well defined average density, independent on
 sample depth as for the simple fractal case.  It can be shown 
 \cite{slp96} that the mass
 function of a MF distribution, in a well defined volume, has
 indeed
 a Press-Schechter behavior.

The
continuous set of exponents
which describes a MF distribution can characterize 
completely  the galaxy distribution when one considers
the mass (or luminosity) of galaxies in the analysis. In 
this way many observational evidences are linked together
and arise naturally from the self-similar properties of the 
distribution.   Considering a MF  distribution, the usual 
power-law space correlation properties correspond just
to  a single exponent of the MF spectrum: such 
an exponent simply describes the space distribution of 
the support of the MF measure.  Furthermore the shape
of the luminosity function (LF), i.e. the probability of 
finding a  galaxy of a certain luminosity per unit volume, 
is related to the MF spectrum of exponents. 
We have shown that, under MF conditions, the LF
is well  approximated by a power law function with an
exponential tail. Such a function corresponds to the 
Schechter LF observed in real galaxy catalogs. In
this case the  shape of the LF is almost independent
on the sample size, 
but  the amplitude of the  LF depends on the sample
 size as a power law function.

These results have important consequences from a 
theoretical point of view. In fact, when one deals 
with self-similar structures the relevant  physical 
phenomenon which leads to the scale-invariant 
structures is determined by the {\it exponent} 
and {\it not the amplitude} of the  physical 
quantities which characterizes such distributions. 

The geometric 
self-similarity has deep implications for the  non-analyticity 
of these structures. In fact, analyticity or regularity would 
imply that at some small scale  the profile becomes smooth 
and one can define  a unique tangent. Clearly this is impossible 
in a self-similar structure because at any small scale a new 
structure appears and the  distribution is never smooth.
 Self-similar structures are therefore intrinsically irregular 
at all scales and correspondingly one  has to change the 
theoretical framework into one which is  capable of 
dealing with non-analytical fluctuations. This means
 going from differential equations to something like 
the  Renormalization Group to study the exponents.
 For example the so-called "Biased theory of galaxy
 formation" \cite{kai84}
is implemented considering 
the evolution of  density fluctuations within an analytic 
Gaussian framework,  while the non-analyticity of fractal 
fluctuations  implies a breakdown of the central limit 
theorem which is the  cornerstone of Gaussian processes 
\cite{slmp98}.

 In this scheme {\it the space correlations and the 
luminosity function are then two aspects of the 
same phenomenon, the MF distribution of visible matter}. 
The more complete and direct way to study such a  
distribution, and hence at the same time the space  
and the luminosity properties, is represented by the 
 computation of the MF spectrum of exponents. 
This is the natural objective  of theoretical investigation 
in order  to explain the formation and the distribution 
of galactic structures. In fact, from a theoretical point
 of view one would like  to identify the dynamical 
processes which can lead to such a  MF distribution.

In this perspective, it would be extremely interesting 
to study the distribution of dark matter and to 
determine its correlation exponent.
It could be that dark matter is distributed 
like an homogeneous fluid, having hence $D=3$
even at small scale. In such a way one may save
the usual FRW metric (which needs an homogenous
density to be developed), while a substantial revision
to the models of galaxy formation is required.
On the contrary if dark matter is found to have the 
same distribution of luminous one, than 
a basic revision of the theory must be considered.
In fact, if dark matter is essentially 
associated to luminous matter, then the use of FRW metric is not
justified anymore.  This does not necessarily imply that there is 
no expansion 
or no Big Bang. 
It implies, however, that  these phenomena should be
described by more complex models \cite{slmp98}. 

It is worth to notice that form an observational point of
view there are various arguments  for the proposition that
galaxies are fair tracers of the mass. 
For example no survey since, in 21-cm, infrared, ultraviolet
 or low surface brightness
optical has revealed a void population.  There 
is a straightforward interpretation: the voids are nearly empty
because they contain little mass \cite{pee98}.

\section{Comparison of N-body simulations with real data} 

\begin{figure}
\epsfxsize 10cm
\centerline{\epsfbox{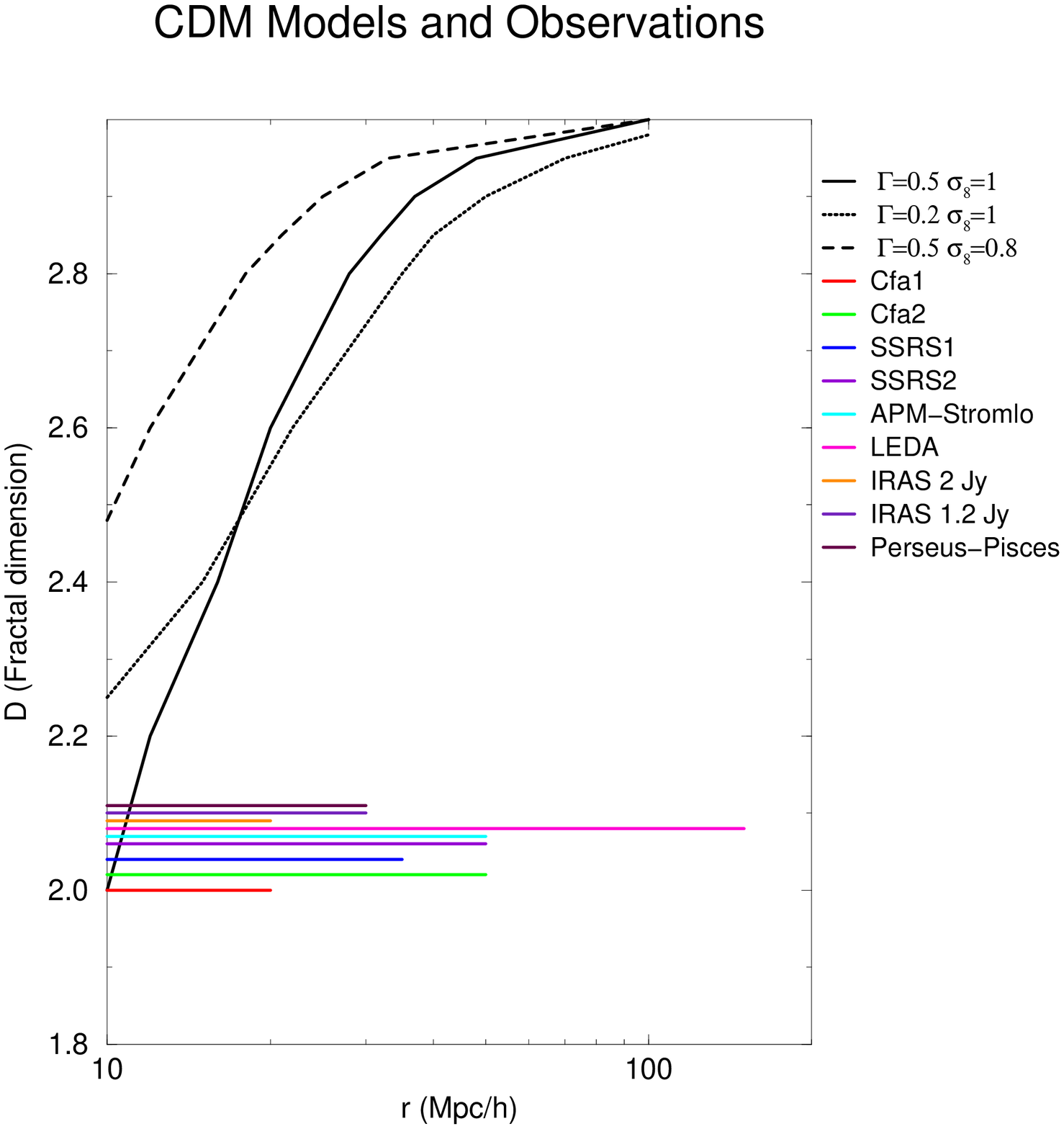}} 
\caption{The fractal dimension versus distance in three
Cold Dark Matter models of power spectra 
with shape and normalized parameters as reported
in the labels (solid lines) (from Wu \etal, 1998).
We show also the different experimental determinations
of the fractal dimension we have obtained.
No agreement can be found at any scale. \label{cdm}}
\end{figure}

 Given the strong evidence for power-law correlations
in the  galaxy  catalogs continuing to scales considerably
larger than the characteristic scale $r_0 = 5\hmp$ for
galaxy clustering which is reproduced by any `adequate'
standard theory of structure formation, the value
of a detailed comparison with the predictions
of such theories is questionable. Further, the evidence in 
all the 3-d catalogs suggest a continuation of 
scale invariance in clustering to
a scale where the contradiction does not require
detailed investigation, and the results 
instead open up a completely new set of problems 
for this field. 
In some recent debate on the issue of homogeneity,
however, it has been suggested that the properties 
of galaxy catalogs observed in the statistics used 
here can in fact be quite compatible with such 
standard theories. For example Wu \etal
\cite{wu98}  have suggested that the 
change in the fractal dimension $D$ as estimated
by various authors is consistent with the 
increase towards the asymptotic value of $D=3$.
In Figure \ref{cdm} we reproduce a figure from this
paper showing the predictions for the ``fractal
dimension'' as a function of scale, as well
as the findings of \cite{slmp98} over the same range
of scale, and the contradiction is manifest. 

In any standard model
of structure formation 
the range we are interested in is that in which 
the non-linear evolution of the perturbations 
is very important, and it is therefore necessary to go
beyond the analysis on which a figure such 
as that of Wu \etal is based.  
We have performed \cite{joyce298} a comparison with the predictions of
standard theories by making use of mock catalogs
compiled by Cole \etal \cite{cole98}, designed 
specifically for comparison with the Sloan Digital Sky Survey.
We thus conclude that our results are in 
disagreement from the predictions for standard 
theories of structure formation.
Further the striking consistency with all the galaxy 3-d
catalogs, some of which probe considerably
large  scales, suggests the
need for a fundamental revision of such theories
(see for a detailed discussion \cite{slmp98,dsl98,bar98}).
In particular the central point of the analysis
is that attention should be shifted 
from correlation amplitudes to correlation exponents, 
as it is only in terms of these concepts that the 
structures seen in red-shift surveys
are correctly characterized.

\section{Conclusions} 
 From the theoretical point of view the fact that 
we have a situation characterized by {\it self-similar structures} 
 implies that we should not use concept
 like $\xi(r)$, $r_0$, $\delta N/N$ and certain properties of
the power spectrum, because they are not suitable to represent 
the real properties of the observed structures. 
To this end also the N-body simulations
should be considered from a new perspective.
One cannot talk about "small" or "large" amplitudes 
for a self-similar structure because of the lack of a reference value like the 
average density.
The Physics should shift from {\it "amplitudes"} towards {\it "exponent" }
and the methods of modern statistical Physics should be adopted.
This requires the development of constructive interactions between two fields.

{\it Possible Crossover.}
We cannot exclude of course, that visible matter 
may really become homogeneous at some large scale not 
yet observed. Even if this would 
happen the best way to identify the
eventual crossover is by 
using the methods we have described and  not the usual ones.
From a theoretical point of view  
the range of fractal fluctuations, extending at least over
three decades ($1 \div 1000 \hmp$), should  anyhow be
addressed with the new theoretical concepts.
Then one should study the (eventual) crossover to homogeneity 
as an additional problem.
For the moment, however, no tendency to 
such a crossover is detectable from the experimental 
data and it may be reasonable to consider also more radical
 theoretical frameworks in which homogenization may 
simply not exist at any scale, at least for  
luminous matter.

{\it Dark Matter.} 
 All our discussion  refers to luminous matter. 
It would be nice if the new picture for the visible universe 
could reduce, to some extent, 
the importance of dark matter in the theoretical framework.
At the moment however this is not clear. We have two possible 
situations: {\it (i)} if dark matter is essentially 
associated to luminous matter, then the use of FRW metric is not
justified anymore.  This does not necessarily imply that there is 
no expansion 
or no Big Bang. 
It implies, however, that  these phenomena should be
described by more complex models\cite{bar98}. {\it (ii)}  If dark matter
 is homogeneous and luminous matter is fractal then, 
at large scale, dark matter  dominate the gravity 
field and the FRW  metric is again valid\cite{bar98}. 
The visible matter however remains self-similar and non analytical 
and it still requires the new theoretical methods 
mentioned before\cite{dsl98}.

\section*{Acknowledgments} 
I would like to thank L. Pietronero
for continuous collaborations.
I am also  in debt  in debt with Y. Baryshev, R. Durrer, 
M. Joyce, 
M. Montuori,
and P. Teerikorpi 
with whom various parts of these work have
been done. 
I warmly thank A. Amici, H. De Vega, 
H. Di Nella, J.P. Eckmann, 
A. Gabrielli, B.B. Mandelbrot, 
D. Pfenniger, 
N. Sanchez and F. Vernizzi 
for useful discussions and collaborations.
Finally I thank the organizers 
for the invitation to this stimulating meeting.
This work has been partially supported by the 
EEC TMR Network  "Fractal structures and  
self-organization"  
\mbox{ERBFMRXCT980183} and by the Swiss NSF.

\end{document}